\documentclass[12pt]{article}
\usepackage{epsf}
\usepackage{amsmath}
\usepackage{graphics}
\usepackage{cite}

\renewcommand{\thefootnote}{\#\arabic{footnote}}
\begin{document}

\newcommand{\gtrsim}{ \mathop{}_{\textstyle \sim}^{\textstyle >} }
\newcommand{\lesssim}{ \mathop{}_{\textstyle \sim}^{\textstyle <} }

\newcommand{\rem}[1]{{\bf #1}}

\renewcommand{\thefootnote}{\fnsymbol{footnote}}
\setcounter{footnote}{0}
\begin{titlepage}

\def\thefootnote{\fnsymbol{footnote}}

\begin{center}
\hfill hep-th/0610213\\
\hfill November 2006\\
\vskip .5in
\bigskip
\bigskip
{\Large \bf Turnaround in Cyclic Cosmology}

\vskip .45in

{\bf Lauris Baum and Paul H. Frampton}

{\em Department of Physics and Astronomy,}

{\em University of North Carolina at Chapel Hill, NC 27599-3255, USA}

\end{center}

\vskip .4in
\begin{abstract}
It is speculated how dark energy in a brane world can help 
reconcile an infinitely cyclic cosmology with the second 
law of thermodynamics. A cyclic model is described, 
in which dark energy with $w<-1$ equation of state
leads to a turnaround at a time, extremely shortly before
the would-be Big Rip, at which both volume and entropy
of our universe decrease by a gigantic factor, while 
very many independent similarly small contracting universes are spawned.
The entropy of our model decreases almost to zero at turnaround
but increases for the remainder of
the cycle by a vanishingly small amount during contraction, empty of 
matter, then by a large factor during inflationary
expansion.

\end{abstract}
\end{titlepage}

\renewcommand{\thepage}{\arabic{page}}
\setcounter{page}{1}
\renewcommand{\thefootnote}{\#\arabic{footnote}}

\newpage

One of the oldest questions in theoretical cosmology is whether
an infinitely oscillatory universe which avoids an initial singularity
can be consistently constructed. As realized by Friedmann\cite{Friedmann}
and especially by Tolman\cite{Tolman,TolmanBook} one principal
obstacle is the second law of thermodynamics which dictates that
the entropy increases from cycle to cycle. If the cycles 
thereby become longer, extrapolation into the past
will lead back to an initial singularity again, thus removing the
motivation to consider an oscillatory universe in the first place.
This led to the abandonment of the oscillatory universe by
the majority of workers.

Nevertheless, an infinitely oscillatory universe is 
a very attractive alternative
to the Big Bang. One new ingredient in the cosmic make-up
is the dark energy discovered only in 1998 and so
it natural to ask whether this can avoid the difficulties with entropy.

Some work has been started to exploit the dark energy
in allowing cyclicity possibly without 
the need for inflation in
\cite{Ekpyrotic,SteinhardtTurok,SteinhardtTurok2,Boyle,SteinhardtTurok3}.
Another new ingredient is the use of branes and a fourth spatial
dimension as in \cite{Sundrum,Randall,Binetruy,Freese} which
examined consequences for cosmology. The Big Rip
and replacement of dark energy by modified gravity were
explored in \cite{PHFTT,PHFTT2}.

If the dark energy has a super-negative equation of state,
$\omega_{\Lambda} = p_{\Lambda}/\rho_{\Lambda} < -1$, it leads
to a Big Rip\cite{caldwell} at a finite future time where there
exist extraordinary conditions with regard to
density and causality as one approaches the Rip.
In the present article we explore whether these exceptional
conditions can assist in providing an infinitely cyclic model. 

We consider a model where, as we approach the Rip, 
expansion stops 
due to a brane contribution
just short of the Big Rip and there is a turnaround
at time $t=t_T$ when the scale factor is deflated to 
a very tiny fraction ($f$) 
of itself and only one causal patch is retained, while
the other $1/f^3$ patches contract independently to
separate universes.
Turnaround takes
place an extremely short time ($< 10^{-27} s$) before the Big Rip would
have occurred, at a time when the universe is 
fractionated into many independent
causal patches \cite{PHFTT2}. 

We discuss contraction which
occurs with a very much smaller universe than in
expansion and with almost vanishing entropy
because it is assumed empty of dust, matter
and black holes.

A bounce takes place 
a short time before a would-be Big Bang. After the bounce, entropy
is injected by inflation\cite{Guth}, where
is assumed that an inflaton field is excited.
Inflation is thus be a part of the present model which is
one distinction from the work of 
\cite{SteinhardtTurok,SteinhardtTurok2,Boyle,SteinhardtTurok3}.  
For cyclicity of the entropy, $S(t) = S(t + \tau)$ to be consistent with thermodynamics 
it is necessary that the deflationary decrease by $f^3$ 
compensate the entropy increase acquired during
expansion including the increase
during inflation.

A possible shortcoming of the proposal could have been the persistence of 
spacetime singularities in cyclic cosmologies\cite{Vilenkin}, but 
to our understanding for the model we outline this problem
is avoided, provided that the time average of the Hubble parameter
during expansion is equal in magnitude and opposite in sign to its average during
contraction.

This model is published because
it gives renewed hope for the infinitely oscillatory universe 
saught in \cite{Friedmann,Tolman,TolmanBook}. Time will 
tell whether the present model is consistent, but at present
we see no fatal flaw.

\bigskip

\noindent {\it Friedmann equation for expansion phase}.
Let the period of the Universe be designated by $\tau$
and the bounce take place at $t = 0$ 
and turnaround at $t = t_T$.
Thus the expansion phase is for times $0 < t < t_T$ 
and the contraction phase corresponds to times
$t_T < t < \tau$.
We employ the following Friedmann equation for
the {\it expansion} period $0 < t < t_T$: 
\begin{equation}
\left( \frac{\dot{a}(t)}{a(t)} \right)^2  =   
\frac{8 \pi G}{3} \left[ \left( \frac{(\rho_{\Lambda})_0}{a(t)^{3(\omega_{\Lambda} + 1)}}
+\frac{(\rho_{m})_0}{a(t)^{3}} +\frac{(\rho_{r})_0}{a(t)^{4}}
\right)
-\frac{\rho_{total}(t)^2}{\rho_c}
\right] 
\label{Friedmann}
\end{equation}
where the scale factor is normalized to $a(t_0)=1$ at the present time
$t = t_0 \simeq 14Gy$. 
To explain the notation, $(\rho_i)_0$ denotes the value of the density $\rho_i$
at time $t=t_0$. The first two terms are the dark energy and total matter
(dark plus luminous) satisfying
\begin{equation}
\Omega_{\Lambda} = \frac{8 \pi G (\rho_{\Lambda})_0}{3 H_0^2} = 0.72 
~~~ {\rm and} ~~~
\Omega_{m} = \frac{8 \pi G (\rho_m)_0}{3 H_0^2} = 0.28 
\end{equation}
where $H_0 = \dot{a}(t_0)/a(t_0)$. The third term in the Friedmann equation
is the radiation density which is now 
$\Omega_r = 1.3 \times 10^{-4}$. 
The final
term $\sim \rho_{total}(t)^2$ is derivable from a
brane set-up\cite{Sundrum,Randall,Freese}; we use
a negative sign arising from negative brane tension (a negative sign
can arise also from a second timelike dimension but that gives 
difficulties with closed timelike paths).
$\rho_{total} = \Sigma _{i=\Lambda, m, r} \rho_{i}$.
As the turnaround is approached, the only significant terms 
in Eq.(\ref{Friedmann}) are the
first (where $\omega_{\Lambda} < -1$) and the last. As the bounce is approached,
the only important terms in Eq.(\ref{Friedmann}) are the third and the last.
(We shall later argue that the second term, for matter, is absent during
contraction.)
In particular, the final term of Eq. (\ref{Friedmann}), $\sim \rho_{total}(t)^2$, 
arising from the brane set-up is insignificant
for almost the entire cycle but becomes dominant
as one approaches $t \rightarrow t_T$ for the turnaround
and again for $t\rightarrow \tau$ approaching the bounce.

\bigskip

\noindent {\it Turnaround}.
Let us assume for algebraic simplicity
$\omega_{\Lambda} = -4/3 = {\rm constant}$. This
value is already almost excluded by WMAP3 \cite{WMAP3} but
to begin we are aiming only at
consistency of infinite cyclicity. More realistic values
may be discussed elsewhere.
With the value $\omega_{\Lambda}=-4/3$
we learn from \cite{PHFTT} that the time to the Big Rip
is $(t_{rip}-t_0) = 11 {\rm Gy} (-\omega_{\Lambda} - 1)^{-1} = 33 {\rm Gy}$
which is, as we shall discuss, within $10^{-27}$ second or less,  when turnaround occurs at $t=t_T$.
So if we adopt $t_0=14 Gy$ then $t_T = t_0 + (t_{rip}-t_0) \sim (14 + 33) Gy = 47 Gy$. 
From the analysis in \cite{PHFTT,PHFTT2,caldwell} 
the time when a system becomes gravitationally unbound corresponds approximately
to the time when the dark energy density matches the mean density
of the bound system. For an object like the Earth or a hydrogen atom 
water density $\rho_{H_2O}$ is a practical unit.

With this in mind, for the simple case of $\omega=-4/3$ we see from
Eq.(\ref{Friedmann}) that the dark energy density grows proportional
to the scale factor $\rho_{\Lambda}(t) \propto a(t)$ and so given
that the dark energy at present is $\rho_{\Lambda} \sim 10^{-29} g/cm^3$
it follows that $\rho_{\Lambda}(t_{H_2O}) = \rho_{H_2O}$ when $a(t_{H_2O}) \sim 10^{29}$. 
We can estimate the time $t_{H_2O}$
by taking on the RHS of the Friedmann equation only dark energy
$\left( \frac{\dot{a}}{a} \right)^2 = H_0^2 \Omega_{\Lambda} a^{-\beta}$
with $\beta=3(1+\omega)$. 
When we specialize to $\omega= - 4/3$ it follows that 
\begin{equation}
\frac{a(t_{H_2O})}{(a(t_0)=1)} = \left( \frac{(t_{rip}-t_0)}
{(t_{rip}-t_{H_2O})}
\right)^{2}
\label{aUa0}
\end{equation}
so that $(t_{rip}-t_{H_2O}) = 33Gy \times 10^{-14.5}
\simeq 10^{3.5}s \sim 1$ hour.  [The value is sensitive to $\omega$]
It is instructive to consider approach to the Rip
a more general critical density $\rho_c = \eta \rho_{H_2O}$ and to
compute the time $(t_{rip} - t_{\eta})$ such that
$\rho_{\Lambda} (t_{\eta}) = \rho_c =  \eta \rho_{H_2O}$.
We then find, using $a(t_{\eta}) = 10^{29} \eta$, that
\begin{equation}
(t_{rip} - t_{\eta}) = (t_{rip}-t_0) 10^{-14.5} \eta^{-1} \simeq \eta^{-1} {\rm hours}
\label{etaC}
\end{equation}
which is the required result. We shall see $\eta>10^{31}$ so the
time in (\ref{etaC}) is $< 10^{-27} s$. 

To discuss the turnaround analytically we keep only the first
and last terms, the only significant ones, on the RHS of Eq.(\ref{Friedmann}) 
which becomes for the special case $\omega = 4/3$
\begin{equation}
\left( \frac{\dot{a}}{a} \right)^2 = \alpha_1 a - \alpha_2 a^2
\end{equation}
in which
\begin{equation}
\alpha_1 = \frac{8 \pi G}{3} (\rho_{\Lambda})_0
~~~~~ \alpha_2 = \frac{8 \pi G}{3} \frac{(\rho_{\Lambda})_0^2}{\rho_c}
\end{equation}
Writing $a=z^2$ and $z = (\alpha_1/\alpha_2)^{1/2} sin\theta$
gives
\begin{equation}
dt = \frac{2 \sqrt{\alpha_2}}{\alpha_1} \frac{d \theta}{sin^2 \theta}
= \frac{2 \sqrt{\alpha_2}}{\alpha_1} d (- cot \theta)
\end{equation}
Integration then gives for the scale factor
\begin{equation}
a(t) = \left( \frac{\alpha_1}{\alpha_2} \right) sin^2 \theta
= \frac{\rho_c}{(\rho_{\Lambda})_0} \left[ \frac{1}{1+\left(\frac{t_T-t}{C} \right)^2} \right]
\end{equation}
where $C = - ( 3/ 2 \pi G \rho_c)^{1/2}$.
At turnaround $t=t_T$, $a(t_T) = [\rho_C/(\rho_{\lambda})_0] = (a(t))_{max}$.
At the present time $t=t_0$, $a(t_0)=1$ and $sin^2\theta_0=[(\rho_{\Lambda})_0/\rho_C] \ll 1$,
increasing during subsequent expansion to $\theta_T = \pi/4$.

A key ingredient in our model is that at turnaround
$t = t_T$ our universe deflates dramatically
with efffective scale factor $a(t_T)$ shrinking 
before contraction to $\hat{a}(t_T) = f a(t_T)$
where $f < 10^{-28}$. 
This jettisoning of almost all, a fraction $(1-f)$, of the
accumulated entropy is permitted by the exceptional causal
structure of the universe. We shall see later that the parameter $\eta$ 
at turnaround lies  
in the range $\eta = 10^{31}$ to $\eta = 10^{87}$
which implies a dark energy density at turnaround
(Planckian density
of $\rho_{\Lambda} \sim 10^{104}\rho_{H_2O}$ can be avoided)
such that, according
to the Big Rip analysis of \cite{PHFTT,PHFTT2}, all known,
and yet unknown smaller,
bound systems have become unbound and the constituents 
causally disconnected.
Recall that the density of a hydrogen atom is approximately
$\rho_{H_2O}$ and we are reaching a dark energy
density of from 31 to 87 orders of magnitude higher.

According to these estimates, at $t=t_T$ the universe has already fragmented
into an astronomical number ($1/f^3$) of causal patches, each of which independently
contracts as a separate universe leading to an infinite multiverse. 
The entropy at $t=t_T$ is thus
divided between these new contracting universes and our universe retains
only a fraction $f^3$. Since our model universe 
has cycled an infinite number of times, the number
of parallel universes is infinite.

\bigskip
\noindent {\it Friedmann equation for contraction phase.}
The contraction phase for our universe occurs for the 
period $t_T < t < \tau$.
The scale factor for the contraction phase will be denoted by $\hat{a} (t)$
while we use always the same linear time $t$ subject to the
periodicity $t + \tau \equiv t$.
At the turnaround we retain a fraction $f^3$ of the entropy with
$\hat{a}(t_T) = f a(t_T)$ and for the contraction phase the Friedmann equation
is
\begin{equation}
\left( \frac{\dot{\hat{a}}(t)}{\hat{a}(t)} \right)^2  =   
\frac{8 \pi G}{3} \left[ \left( \frac{(\hat{\rho}_{\Lambda})_0}{\hat{a}(t)^{3(\omega_{\Lambda} + 1)}}
+\frac{(\hat{\rho}_{r})_0}{\hat{a}(t)^{4}} \right)
-\frac{\hat{\rho}_{total}(t)^2}{\hat{\rho}_c}
\right] 
\label{hatFriedmann}
\end{equation}
where we have defined
\begin{equation}
\hat{\rho}_i(t) = 
\frac{(\rho_i)_0 f^{3(\omega_i +1)}}{\hat{a}(t)^{3(\omega_i+1)}}  
= \frac{(\hat{\rho}_i)_0}{\hat{a}(t)^{3(\omega_i+1)}}  
\label{hatrho}
\end{equation}
but in contrast to Eq.(\ref{Friedmann}) we have set $\hat{\rho}_m = 0$
because our hypothesis is that the causal patch retained in the model
contains only dark energy
and radiation but no matter including no black holes. 
This is necessary because during a contracting 
phase dust or matter would clump, even more readily than during expansion, and
inevitably interfere with cyclicity. 
Perhaps more importantly, presence of dust or matter would require
that our universe go in reverse through several phase transitions 
(recombination, QCD and electroweak to name a few) which would violate
the second law of thermodynamics.
We thus require that {\it our universe comes back empty!} 
Any tiny entropy associated with radiation is
constant during adiabatic contraction.

The contraction of our universe
will proceed from one of the $1/f^3$ causal patches following
Eq.(\ref{hatFriedmann}) until the radiation balances the
brane tension at the bounce.

\bigskip

\noindent {\it Bounce.}
At the bounce, 
the contraction scale is given, using $\rho_c = \eta \rho_{H_2O}$,
from Eq. (\ref{Friedmann})
as
\begin{equation}
a (\tau)^4 = \left( \frac{ (\rho_r)_0}{\eta \rho_{H_2O}} \right) 
\label{atau}
\end{equation}
Now the model's bounce at $t=\tau$ must be before 
the electroweak transition at
$t_{EW}=10^{-10}s$ when $a(t_{EW}) = 10^{-15}$, and after
the Planck scale when $a(t_{Planck}) = 10^{-32}$ in order to
accommodate the well established weak transition and to
avoid uncertainties associated with quantum gravity.
With this in mind, here are
three illustrative values (A, B, C) for the bounce temperature $T_B$:

\begin{itemize}

\item (A) At a GUT scale $T_B=10^{17} GeV, a(t_B)=10^{-30}$.

\item (B) At an intermediate scale $T_B=10^{10} GeV, a(t_B)=10^{-23}$.

\item (C) At a weak scale $T_B=10^{3} GeV, a(t_B)=10^{-16}$.

\end{itemize}

From Eq.(\ref{atau}) and Eq.(\ref{etaC}) for these three cases
one finds

\begin{itemize}

\item (A) $\eta = 10^{87}$ and $(t_{rip}-t_T) = 10^{-87} hr$.
 
\item (A) $\eta = 10^{59}$ and $(t_{rip}-t_T) = 10^{-59} hr$.

\item (A) $\eta = 10^{31}$ and $(t_{rip}-t_T) = 10^{-31} hr$.

\end{itemize}

Immediately after the bounce, we assume that an
inflaton field is excited and there is conventional
inflation with enhancement E = $a(\tau+\delta)/\hat{a}(\tau)$.
Successful inflation requires $E > 10^{28}$. Consistency
requires therefore $f < E^{-1}$ to allow for the entropy
accrued during expansion after inflation. The fraction
of entropy jettisoned from our universe at deflation 
is thus extremely close to one, being less than one and more than $(1 - 10^{-28})^3$.

\bigskip

\noindent {\it Entropy.}
Consider first the present epoch $t=t_0$. The contributions
of the radiation to the entropy density $s$ follows
the relation
\begin{equation}
s = \frac{2 \pi^2}{45}g_* T^3
\label{entropy}
\end{equation}
Photons contribute $g_*=2$he present
CMB temperature is $T=2.73K \equiv 0.235 meV \sim 1.191 (mm)^{-1}$. 
Substitution in Eq.(\ref{entropy}) gives a present radiation 
entropy density $s_{\gamma}(t_0)= 1.48 (mm)^{-3}$. Using a volume estimate
$V=(4\pi/3)R^3$ with $R=0Gly \simeq \times10^{29}mm$ gives a total
radiation entropy
$S_{\gamma} \sim 6.3 \times 10^{87}$. Including neutrinos increase $g_*$ in
Eq.(\ref{entropy}) from $g_*=2$ to $g_*=3.36=2+6\times(7/8) \times (4/11)^{4/3}$.
This increases $S_{\gamma}=6.3 \times 10^{87}$ to $S_{\gamma+\nu} \sim \times 10^{88}$. 
                                   
This total entropy is interpretable as $exp (10^{88})$ degrees of freedom,
or in information theory\cite{KN} to a number $I$ of qubits where
$2^I = e^S$ so that $I = S/(ln 2 = 0.693) \sim 10^{88}$. This is well below
the holographic bound which is dictated by the area in terms
of Planck units $10^{-64} mm^2$ which gives $S_{holog}(t_0) = 4\pi (10^{29} mm)^2
/(10^{-32} mm)^2 \sim 10^{123}$ about $10^{35}$ times bigger.
In \cite{KN} it is suggested that at least some of this difference may
come from supermassive black holes.
The entropy contribution from the baryons is smaller than $S_{\gamma}$
by some ten orders of magnitude, so like that of the dark matter,
is negligible. 

What is the entropy of the dark energy? If it is perfectly homogeneous
and non-interacting it has zero temperature and entropy. 
Finally, the 4th term in Eq.(\ref{Friedmann}) corresponding
to the brane term is neglible, as we have already estimated. 
The conclusion is that at present $S_{total}(t_0) \sim 10^{88}$.

Now consider the entropy approoaching turnaround at $t=t_T$.
We have estimated that $a(t_T) = 10^{29} \eta$ and representative
values for $\eta = \rho_c/\rho_{H_2O}$ are
$10^{31}, 10^{59}$ and $10^{87}$. The temperature
$T_{\gamma}$ of the radiation scales as $T_{\gamma} \propto a(t)^{-1}$
so using the entropy
density of Eq.(\ref{entropy}) a comoving 3-volume 
$\propto a(t)^3$ will contain
the same total radiation entropy $S_{\gamma}(t_T) = S_{\gamma}(t_0)$
as at present; this is simply the usual adiabatic expansion.
The expansion from $t=0$ to $t_T$ 
is not purely adiabatic
because irreversible processes take place. The first is inflation
which increases entropy by $> 10^{84}$. There are
phase transitions such as  
the electroweak transition at $t_{EW}\sim 100ps$, the
QCD phase transition at $t_{QCD} \sim 100 \mu s$, 
and recombination at $t_{rec}\sim 10^{13}s$. Further
irreversible processes occur during during stellar evolution.
Although the expansion of the radiation, the
dominant contributor to the entropy, is
adiabatic, the entropy of matter
increases in accord with the
second law of thermodynamics.
In our model, the entropy of the matter increases
between $t = 0$ and $t_T \sim 47Gy$. 
Setting the entropy of the dark energy to zero and the
radiation as adiabatic, the matter part represented
by $\rho_m$ will cause the entropy to rise
from $S(t = 0)$ to $S(t_T) = S(t = 0) + \Delta S$ where $\Delta S$
causes the contradiction plaguing previous oscillatory
model universes\cite{Friedmann,Tolman,TolmanBook}.

Our main point is that in order for entropy 
to be cyclic, the entropy which was enhanced by a huge
factor $E^3 > 10^{84}$ at inflation must be 
reduced dramatically at some point during the cycle
so that $S(t) = S(t+\tau)$ becomes possible.  Since it
increases during expansion and contraction, the only
logical possibility is the decrease
at turnaround as accomplished by our causal patch idea. 
The second law of thermodynamics
continues to obtain for other causal patches, each
with practically vanishing entropy at turnaround, but these are 
permanently removed from our universe 
contracting instead into separate universes.

For contraction $t_T < t < \tau$ 
we are assuming the universe during
contraction is empty of matter until the bounce so
its entropy is vanishingly small.
Immediately after the bounce inflation 
arises from an inflaton field, assumed to be excited.
We find the counterpoise of inflation at the bounce
and deflation at turnaround an appealing aspect of
the present model.   

{\it Conclusion.}
The standard cosmology based on a Big Bang augmented by
an inflationary era is impressively consistent with the detailed
data from WMAP3 \cite{WMAP3} when dark energy, most conservatively
a cosmological constant, is included.
Our objections to this standard model
are more aesthetic than motivated directly by observations.
The first objection is the nature of the initial singularity and the initial
conditions. A second objection, not of concern to all colleagues, is that 
the predicted fate of the universe is an infinitely long expansion.
We have outlined here a cyclic cosmology resting on phantom
dark energy where these objections
are ameliorated: the classical density and temperature never become
infinite and future expansion is truncated. 
Also, our proposal of deflation naturally 
leads to a multiverse picture, somewhat reminiscent of
that predicted in eternal inflation, though here
the proliferation of universes must be infinite and
originates at the opposite
end of a cyclic cosmology, at its maximum rather than at its minimum size.

We publish our infinitely cyclic model mainly in the hope that
it will stimulate a more detailed and compelling
formulation.

\bigskip
\bigskip
\bigskip
\bigskip

\begin{center}

{\bf Acknowledgements}

\end{center}

\bigskip
We thank Alex Vilenkin and Robert Wald for useful discussions.
This work was supported in part by the
U.S. Department of Energy under Grant No. DE-FG02-06ER41418.

\end{document}